\documentclass[aps,12pt,eqsecnum]{revtex4}
\usepackage{amsmath}
\usepackage{graphicx}
\newcommand{\beq}{\begin{equation}}
\newcommand{\eeq}{\end{equation}}
\newcommand{\bes}{\begin{subequations}}
\newcommand{\ees}{\end{subequations}}
\newcommand{\bea}{\begin{eqnarray}}
\newcommand{\eea}{\end{eqnarray}}

\begin{document}
\title{Relativistic Coulomb excitation at small impact parameters}
\author{B.F. Bayman\\
{\it School of Physics and Astronomy, University of Minnesota,}\\
{\it 116 Church Street S.E., Minneapolis, MN 55455, U.S.A.}\\
and\\
F. Zardi\\
{\it Istituto Nazionale di Fisica Nucleare and Dipartimento di Fisica, \\}
{\it Via Marzolo,8 I-35131 Padova,Italy.}}

\begin{abstract}
The semiclassical model of relativistic Coulomb excitation is studied in situations in which the impact parameter is small enough so that projectile and target charge distributions overlap. The electromagnetic effects of this overlap are shown to be small. Realistic nucleon-nucleon reaction cross-sections, and realistic nuclear radial charge and matter distributions are used to determine a formula for the lower impact parameter limit to be used in the calculation of the Coulomb excitation cross-section. A wide selection of projectile-target pairs is explored, in the bombarding energy range of 1 GeV to 5 GeV per nucleon.  
\end{abstract}

\maketitle
%\receipt{\today}
\date{\today}
\section{Introduction}
The semi-classical approach to relativistic Coulomb excitation (RCE) assumes that the Coulomb impulse suffered by the projectile nucleus as it passes the target nucleus is small compared to the linear momentum of the projectile. Then the deflection of the projectile will be small, and its trajectory can be approximated by a straight line, with impact parameter $b$. If $b$ is large enough so that the projectile and target do not experience each others nuclear forces, the only processes that can occur are electromagnetic. For example, as the projectile moves along its trajectory, the electromagnetic fields it produces can induce induce transitions in the target\footnote{Transitions can also be induced in the projectile, due to the electromagnetic fields of the target. In this paper, we will assume that the projectile remains in its ground state throughout the collision.}. The RCE cross-section for the population of a final target state $\psi_\alpha$, starting with the target ground state $\psi_{{\rm g.s.}}$, is given by
\beq
\sigma_{{{\rm g.s.}}  \rightarrow \alpha}=2 \pi \int_{b=b_{{\rm min}}}^\infty
~b ~db ~P_{{{\rm g.s.}}\rightarrow \alpha}(b),
\label{Eqn1.1}
\eeq
where $P_{{{\rm g.s.}} \rightarrow \alpha}(b)$ is the probability of the $\psi_{{\rm g.s.}} \rightarrow\psi_\alpha$ target transition when the target experiences the electromagnetic impuse due to the projectile following an orbit with impact parameter $b$.

The choice of $b_{{\rm min}}$ can have a significant effect on $\sigma_{{\rm g.s.}\rightarrow\alpha}$ calculated with Equation (\ref{Eqn1.1}), since $P_{{\rm g.s.} \rightarrow \alpha}(b)$ has its greatest value at $b=b_{{\rm min}}$. This is because the projectile electromagnetic fields at the target are greatest when the two are in closest proximity. Also, for small $b$ the electromagnetic impulse at the target is more sudden, which makes it more effective at exciting high-energy target states, such as giant multipole excitations. One of the goals of this paper is to provide reliable values of $b_{{\rm min}}$ for a wide variety of projectile-target pairs and a wide range of relativistic bombarding energies.

The assumption behind Equation (\ref{Eqn1.1}) is that there are no nuclear interactions when $b \geq b_{{\rm min}}$, but when $b<b_{{\rm min}}$ the nuclear interactions are so strong that they dominate over electromagnetic processes, and we no longer have Coulomb excitation. We will refer to this latter situation using the term {\it absorption}. A more realistic discussion would involve a continuous transition as $b$ increases, from complete absorption at small $b$, to zero absorption at large $b$. If $X(b)$ is the probability that a projectile traverses its orbit with no nuclear interaction with the target, the absorption probability would be $1-X(b)$, and the RCE cross-section would be   
\beq
\sigma_{{\rm g.s}\rightarrow \alpha}=2 \pi \int_{b=0 }^\infty~b ~db ~P_{{{\rm g.s.}}\rightarrow \alpha}(b) \times X(b),
\label{Eqn1.2}
\eeq 
The function $X(b)$ can be estimated by folding the nuclear density distributions of the projectile and target with the nucleon-nucleon interaction cross-section.

For the small-$b$ range of the integral in Equation (\ref{Eqn1.2}) there are situations in which the tails of the projectile and target overlap. When this happens, the electromagnetic interaction between the projectile and target is more complicated than the situation of Equation  (\ref{Eqn1.1}), where all the projectile charge density is assumed to be outside the target.
In this connection, we explore the $b$-range of the integral in Equation (\ref{Eqn1.2}) where $X(b)$ is making the transition from a small value (almost complete absorption) to near unity (almost no absorption). 

More precisely, in Sec.II we consider the nucleon-nucleon interactions, and an
explicit form will be deduced for the function $X(b)$.
In Sec.III we will investigate the effect on $P_{{\rm g.s.}\rightarrow \alpha}(b)$ of the overlapping of the projectile and target charge densities. We will then discuss in Sec.IV several prescriptions for the parameter $b_{{\rm min}}$ of Equation (\ref{Eqn1.1}) which will lead to the same calculated cross-section as the more accurate Equation (\ref{Eqn1.2}). 

\section{Nuclear interaction between projectile and target}

We discuss the nuclear interactions between the projectile and target nuclei in terms of the optical limit of the Glauber model \cite{Gla}, in which each projectile nucleon is treated as a grey disk, characterized by a {\it profile function} $\gamma(s)$. In Figure 1, the projectile and target centers are labelled $O_{_{\rm P}}$ and $O_{_{\rm T}}$, respectively. The edge view of a disk is shown, representing a projectile nucleon located at ${\bf r}_{{\rm P}}$ relative to $O_{{\rm P}}$. Points in the plane of the disk are located by ${{\bf s}}$ relative to the center of the disk (the center of the projectile nucleon). If the projectile moves a distance $dz$, then the disk sweeps out an effective reaction volume of 
\beq
dz~\int d^2s~\gamma(s),
\label{Eqn2.1}
\eeq
where $\gamma(s)$ is the profile function, related to the nucleon-nucleon scattering amplitude $f_{{\rm NN}}({\bf q})$ by a two-dimensional Fourier transform
\beq
\gamma({\bf s})=\frac{1}{2 \pi i k_{_{\rm NN}}} \int d^2 q~e^{-i {\bf q} \cdot {\bf s}}f_{_{\rm NN}}({\bf q}).
\label{Eqn2.2}
\eeq
If $~n_{_{\rm T}}({\bf r}_{_{\rm T}})$ is the number density of target nucleons, then the probability that a target nucleon is in the effective reaction volume given by Equation(\ref{Eqn2.1}) is
$$
dz~ \int d^2 s ~\gamma(s)~n_{_{\rm T}}({\bf r}_{_{\rm T}}),
$$
where
\beq
{\bf r}_{_{\rm T}}~=~{\bf b}~+~z{\bf {\hat z}}~+~{\bf r}_{_{\rm P}}~+~{\bf s}.
\label{Eqn2.3}
\eeq

This refers to a single projectile nucleon. If $~n_{_{\rm P}}({\bf r}_{_{\rm P}})$ is the number density of projectile nucleons, the total probability of a nuclear interaction when the projectile moves a distance $dz$ is
$$
dz~\int d^3 r_{_{\rm P}}~n_{_{\rm P}}({\bf r}_{_{\rm P}})~\int d^2 s ~\gamma(s)~n_{_{\rm T}}({\bf r}_{_{\rm T}})~\equiv~dz~f(z).
$$
Therefore the probability $P(b)$ that {\it no} nuclear interactions occur when $z$ goes from $-\infty$ to $\infty$ is
\beq
X(b)=e^{-\int_{-\infty}^{\infty}~f(z)~dz}~=~e^{-\int_{-\infty}^{\infty}dz\int d^3 r_{_{\rm P}} ~n_{_{\rm P}}({\bf r}_{{\rm P}})~\int d^2 s~\gamma(s)~n_{_{\rm T}}({\bf r}_{_{\rm T}})}
\label{Eqn2.4}
\eeq

Equation (\ref{Eqn2.3}) implies that, for fixed ${\bf r}_{_{\rm P}}$ and $b$, 
\begin{eqnarray*}
d^2 s~dz~&=&~d^2({\bf r}_{_{\rm T}})_\perp~dz_{_{\rm T}}~=~d^3~r_{_{\rm T}}\\
{\bf s} &=& ({\bf r}_{_{\rm T}})_\perp~-({\bf r}_{_{\rm P}})_\perp~-~{\bf b}
\end{eqnarray*}
Thus Equation (\ref{Eqn2.4}) can be re-written as 
\beq
X(b)~=~e^{ -\int d^3 r_{_{\rm P}}~d^3 r_{_{\rm T}}~n_{_{\rm P}}({\bf r}_{_{\rm P}})n_{_{\rm T}}({\bf r}_{_{\rm T}})~\gamma(|({\bf r}_{_{\rm T}})_\perp-({\bf r}_{_{\rm P}})_\perp-{\bf b}|)}
\label{Eqn2.5}
\eeq
This is the $X(b)$ needed in Equation (\ref{Eqn1.2}).

To evaluate the integral in the exponential of Equation (\ref{Eqn2.5}), we express the number densities in terms of their Fourier transforms. For spherically symmetric nuclei we define
\bes
\bea
{\tilde n}_{_{\rm P}}({\bf q}) &\equiv&\int d^3 r_{_{\rm P}}~n_{_{\rm P}}({\bf r}_{_{\rm P}})e^{i {\bf q}\cdot{\bf r}_{_{\rm P}}}=\frac{4\pi}{q}\int_0^\infty r \sin(qr)n_{_{\rm P}}(r)dr ={\tilde n}_{{\rm P}}(q) \label{Eqn2.6a}\\
{\tilde n}_{_{\rm T}}({\bf q})&\equiv&\int d^3 r_{_{\rm T}}~n_{_{\rm T}}({\bf r}_{_{\rm T}})e^{i {\bf q}\cdot{\bf r}_{_{\rm T}}}=\frac{4\pi}{q}\int_0^\infty r \sin(qr)n_{_{\rm T}}(r)dr ={\tilde n}_{_{\rm T}}(q) \label{Eqn2.6b}
\eea
\ees
With the help of Equation (\ref{Eqn2.2}), we can write
$$
 \int d^3 r_{_{\rm P}}~d^3 r_{_{\rm T}}~n_{_{\rm P}}({\bf r}_{_{\rm P}})n_{_{\rm T}}({\bf r}_{_{\rm T}})~\gamma(|({\bf r}_{_{\rm T}})_\perp-({\bf r}_{_{\rm P}})_\perp-{\bf b}|)
$$
$$
=\frac{1}{2 \pi i k_{_{\rm NN}}}\int d^2q_\perp~e^{i {\bf q}_\perp \cdot {\bf b}}~ {\tilde n}_{_{\rm P}}(q_\perp) {\tilde n}_{_{\rm T}}(q_\perp)~f_{_{\rm NN}}(q_\perp)
$$
\beq
=\frac{1}{ik_{_{\rm NN}}}\int_0^\infty q_\perp dq_\perp J_0(q_\perp b)~{\tilde n}_{_{\rm P}}(q_\perp) {\tilde n}_{_{\rm T}}(q_\perp)~f_{_{\rm NN}}(q_\perp)
\label{Eqn2.7}
\eeq

We have so far not distinguished between neutrons and protons. Since scattering amplitudes for identical and non-identical nucleons are different, the integral in Equation (\ref{Eqn2.5}) should be replaced by
$$
\int d^3 r_{_{\rm P}}~d^3 r_{_{\rm T}}~ \left[n_{_{\rm P,Z}}({\bf r}_{_{\rm P}})n_{_{\rm T,Z}}({\bf r}_{_{\rm T}})+n_{_{\rm P,N}}({\bf r}_{_{\rm P}})n_{_{\rm T,N}}({\bf r}_{_{\rm T}})\right]~\gamma_{{\rm ZZ}}(|({\bf r}_{_{\rm T}})_\perp-({\bf r}_{_{\rm P}})_\perp-{\bf b}|)
$$
$$
+~\int d^3 r_{_{\rm P}}~d^3 r_{_{\rm T}}~ \left[n_{_{\rm P,Z}}({\bf r}_{_{\rm P}})n_{_{\rm T,N}}({\bf r}_{_{\rm T}})+n_{_{\rm P,N}}({\bf r}_{_{\rm P}})n_{_{\rm T,Z}}({\bf r}_{_{\rm T}})\right]~\gamma_{{\rm ZN}}(|({\bf r}_{_{\rm T}})_\perp-({\bf r}_{_{\rm P}})_\perp-{\bf b}|)
$$
Here the subscripts $Z$ and $N$ distinguish between protons and neutrons, whether occurring in the projectile or target. Similarly, Equation (\ref{Eqn2.2}) should be replaced by
\begin{eqnarray*}\gamma_{{\rm ZZ}}({\bf s})&=&\frac{1}{2 \pi i k_{{\rm ZZ}}} \int d^2 q~e^{-i {\bf q} \cdot {\bf s}}f_{{\rm ZZ}}({\bf q})\\
\gamma_{{\rm ZN}}({\bf s})&=&\frac{1}{2 \pi i k_{{\rm ZN}}} \int d^2 q~e^{-i {\bf q} \cdot {\bf s}}f_{{\rm ZN}}({\bf q}).
\end{eqnarray*}

If the center-of-mass energy of the projectile and target nucleons is in the GeV region, as in relativistic Coulomb excitation, the spin-averaged nucleon-nucleon scattering amplitudes can be parameterized by (see, e.g., ref.\cite{Igo})
\begin{eqnarray*}
\frac{f_{{\rm ZZ}}(q_\perp)}{i k_{{\rm ZZ}}}&\sim& \frac{\sigma_{{\rm ZZ}}}{4 \pi}~e^{-B_{{\rm ZZ}}(\hbar q_\perp)^2}\\
\frac{f_{{\rm ZN}}(q_\perp)}{i k_{{\rm ZN}}}&\sim& \frac{\sigma_{{\rm ZN}}}{4 \pi}~e^{-B_{{\rm ZN}}(\hbar q_\perp)^2}
\end{eqnarray*}
The numerical values of $\sigma_{{\rm ZZ}},~B_{{\rm ZZ}}, \sigma_{{\rm ZN}}$ and $B_{{\rm ZN}}$, which are functions of energy, were taken from the article by Igo \cite{Igo}. They are given in Table 1.

The nucleon number densities required in Equations (\ref{Eqn2.6a},\ref{Eqn2.6b}) were taken from the IAEA compilation \cite{IAEA}. They are the results of Hartree-Fock-Bogoliubov calculations, whose parameters are fitted to measured nuclear masses. In a comparison with 523 nuclei, these number densities agreed with measured radii within an rms error of .028 fm. 

Bertulani {\it et al} \cite{BCH} have used a similar approach to represent the effect of nuclear  absorption in grazing collisions. Their analysis assumes that the profile function $\gamma(s)$ has zero extension, equivalent to the assumption that the nucleon-nucleon scattering amplitude is ${\bf q}$ independent. They do not distinguish between the interaction of identical and non-identical nucleons. In different contexts, a similar approach has been developed by Benesh,
Cook and Vary \cite{BCV} and by Kox {\it et al} \cite{KGP}

\section{The electromagnetic transition matrix elements with overlapping charge distributions}
In the previous Sections, we saw that at the bottom of the range of impact parameters there can be overlapping of the projectile and target charge distributions. We now consider whether this requires us to modify the standard analysis of relativistic Coulomb excitation, which is based on the assumption that the projectile and target charge distributions do not overlap.

\subsection{The scalar and vector potentials}
We orient our axes so that the projectile center moves with velocity $v{\bf {\hat z}}$ relative to the target center in their common ${\bf {\hat y}}-{\bf {\hat z}}$ plane. If $(x,y,z,t)$ locate an event relative to the projectile, and $(x',y',z',t')$ locate the same event relative to the target, then
$$
x=x',~~~y=y'-b,~~~z=\gamma(z'-vt'),~~~t=\gamma(t'-\frac{v}{c^2}z'),~~~\gamma\equiv(1-\frac{v^2}{c^2})^{-\frac{1}{2}}
$$ 

Suppose that the projectile charge density is static and spherically symmetric in its own rest frame, so that it can be written
$$
\rho_{_{\rm P}} (x,y,z)=\rho_{_{\rm P}} (r),~~~r\equiv\sqrt{x^2+y^2+z^2}.
$$
Then the scalar potential due to this charge distribution, as measured in the projectile rest frame, is given by
\beq
\phi_{_{\rm P}}(r)~=~\frac{4\pi}{r}\int_{s=0}^r~s^2\rho_{_{\rm P}}(s)ds
+4\pi\int_{s=r}^\infty~s\rho_{_{\rm P}}(s)ds
\label{Eqn3.1}
\eeq
It can be readily verified that this expression for $\phi_{_{\rm P}}(r)$ satisfies Poisson's equation for the specified projectile charge density:
$$
\nabla^2_{_{\bf{\rm r}}}\phi_{_{\rm P}}(r)~=~\frac{1}{r^2}\frac{d}{dr}r^2\frac{d}{dr}~
\phi_{_{\rm P}}(r)~=~-4 \pi \rho_{_{\rm P}} (r).
$$

The scalar potential measured by an observer at the target center can be determined from Equation(\ref{Eqn3.1}), since the scalar potential transforms as the 0-component of a Lorentz 4-vector. This leads to 
\bea
\phi_{_{\rm P}}'(x',y',z',t')&=&\gamma\phi_{_{\rm P}}(x,y,z,t)\nonumber \\
~&=&\frac{4\pi\gamma}{r}\int_{s=0}^r~s^2\rho_{_{\rm P}}(s)ds
+4\pi\gamma\int_{s=r}^\infty s\rho_{_{\rm P}}(s)ds
\label{Eqn3.2}
\eea
with $r=\sqrt{x'^2+(y'-b)^2+\gamma^2(z'-vt')^2}$. Note that there is no contribution in Equation(\ref{Eqn3.2}) from a projectile current density, as seen by a projectile-based observer. This would be the situation, for example, if the projectile were a doubly-closed-shell nucleus, in which every occupied shell-model state was matched by an occupied time-reversed state.  

It is helpful to modify Equation(\ref{Eqn3.2}) by adding and subtracting the quantity $\frac{4 \pi \gamma}{r}\int_r^\infty s^2\rho_{_{\rm P}} (s)ds$. This converts Equation(\ref{Eqn3.2}) into
\bes
\beq
\phi_{_{\rm P}}'(x',y',z',t')=\frac{4 \pi \gamma}{r}\int_{s=0}^\infty s^2\rho_{_{\rm P}} (s)ds ~+~4\pi\gamma\int_{s=r}^\infty (s-\frac{s^2}{r})\rho_{_{\rm P}} (s)ds.
\label{Eqn3.3a}
\eeq
Since the total projectile charge is $Z_{_{\rm P}}e=\int_{s=0}^\infty 4\pi s^2\rho_{_{\rm P}} (s)ds$, this equation can be re-written as
\beq
\phi_{_{\rm P}}'(x',y',z',t')=\frac{\gamma Z_{_{\rm P}}e}{\sqrt{x'^2+(y'-b)^2+\gamma^2(z'-vt')^2}}~+~4\pi\gamma\int_{s=r}^\infty (s-\frac{s^2}{r})\rho_{_{\rm P}} (s)ds.
\label{Eqn3.3b}
\eeq
\ees

The first term on the right-hand sides of Equations(\ref{Eqn3.3a},\ref{Eqn3.3b}) is the Lienard-Wiechert (LW) potential (see, e.g., \cite{Jack}), which is the potential at $(x',y',z',t')$ produced by a {\it point charge} $Z_{_{\rm P}}e$ at the projectile center, as measured by a target-based observer. The second term is a correction to the LW potential, needed when the projectile has charge that extends farther from the projectile center than the observation point $(x',y',z',t')$. In other words, this correction term is needed for points in space within the projectile charge distribution. It is seen that although the LW term involves only the total projectile charge $Z_{_{\rm P}}e$, the overlap correction term depends upon the radial shape of the projectile charge distribution.

The same Lorentz transformation that yields the scalar potential (\ref{Eqn3.2}) yields the vector potential
\beq
{\bf A}_{_{{\rm P}}}'({\bf r'},t')~=~\frac{v}{c}~\phi_{_{{\rm P}}}'({\bf r'},t')~{\bf {\hat z}},
\label{Eqn3.4}
\eeq

\subsection{Calculation of the transition charge density}
Let us consider the specific case of a direct transition between the zero- and one-phonon states of the target giant dipole excitation. We will generate the transition charge and current densities with the Goldhaber-Teller model 
\cite{GTel} of the giant dipole resonance (GDR), in which the target protons and neutrons oscillate relative to each other. For example, phonon excitations of the $^{40}$Ca ground state can be constructed in which spherical clusters of 20 protons and 20 neutrons oscillate relative to each other, with the oscillation degree of freedom the vector ${\bf r}_{{\rm pn}}$, drawn from the center of the neutron cluster to the center of the proton cluster (see Figure 2). If the oscillations are small, the GDR potential can be expected to be approximately harmonic, and then the relative motion of the cluster centers would be governed by a harmonic oscillator wave function $\psi^{n,\ell}_m({\bf r}_{_{\rm pn}})$. The ground state relative motion would be  determined by $\psi^{0,0}_0({\bf r}_{_{\rm pn}})$, and the ground state charge density would be
\bes
\beq
\rho_{{\rm g.s.}}({\bf r'})=e\int d^3r_{_{{\rm pn}}}~\left[\psi^{0,0}_{0}({\bf r}_{_{\rm pn}})\right]^*~\psi^{0,0}_{0}({\bf r}_{_{\rm pn}})~G\left(\left|{\bf r'}-f {\bf r}_{{\rm pn}} \right|\right),
\label{Eqn3.5a}
\eeq 
where $G(s)$ is the number density of the protons at a distance $s$ from the center of the protons (see Figure 2), and $f$ is defined by
\beq
f \equiv \frac{N_{{\rm T}}}{Z_{{\rm T}}+N_{{\rm T}}}=\frac{N_{{\rm T}}}{A_{{\rm T}}},
\label{Eqn3.5b}
\eeq
so that $f {\bf r}_{{\rm pn}}$ is the vector connecting the target center of mass to the center of the target proton cluster, which we take to be an inert sphere. The transition charge density for the transition from the ground state to the GDR state with relative motion $(n,\ell,m)$ is given by
\beq
\rho_{{\rm g.s.}\rightarrow (n,\ell,m)}({\bf r'})=e\int d^3r_{_{{\rm pn}}}~\left[\psi^{n,\ell}_{m}({\bf r}_{_{\rm pn}})\right]^*~\psi^{0,0}_{0}({\bf r}_{_{\rm pn}})~G\left(\left|{\bf r'}-f {\bf r}_{{\rm pn}} \right|\right),
\label{Eqn3.5c}
\eeq 
\ees

It will be sufficient for our purposes to consider the transition from the ground state to the mode 
$$
\psi_{{\rm y}}~\equiv~\frac{1}{\sqrt{2}}~\left[~\psi^{0,1}_1~+~\psi^{0,1}_{-1}~\right],
$$
which is one quantum of oscillation in the ${\bf {\hat y}}$ direction. At high bombarding energy, this is the strongest direct transition. Because this state has a simple interpretation in a Cartesian representation, we re-write Equations (\ref{Eqn3.5a}, \ref{Eqn3.5c}) in terms of one-dimensional harmonic oscillator eigenstates:
\bes
\begin{eqnarray}
\rho_{{\rm g.s.}}({\bf r'})&=&e\int dx_{{\rm pn}}dy_{{\rm pn}}dz_{{\rm pn}}\left[\psi_0(x_{{\rm pn}})\right]^2\left[\psi_0(y_{{\rm pn}})\right]^2\left[\psi_0(z_{{\rm pn}})\right]^2\nonumber \\
&\times&G\left(\sqrt{\left(x'-f~x_{{\rm pn}}\right)^2+\left(y'-f~y_{{\rm pn}}\right)^2+\left(z'-f~z_{{\rm pn}}\right)^2}\right) \label{Eqn3.6a}\\
\rho_{ {{\rm g.s.}}\rightarrow{{\rm y}}}({\bf r'})&=&e\int dx_{{\rm pn}}dy_{{\rm pn}}dz_{{\rm pn}}\left[\psi_0(x_{{\rm pn}})\right]^2\left[\psi_1(y_{{\rm pn}})\psi_0(y_{{\rm pn}})\right]\left[\psi_0(z_{{\rm pn}})\right]^2\nonumber \\
&\times&G\left(\sqrt{\left(x'-f~x_{{\rm pn}}\right)^2+\left(y'-f~y_{{\rm pn}}\right)^2+\left(z'-f~z_{{\rm pn}}\right)^2}\right) \label{Eqn3.6b}
\end{eqnarray}
\ees   
The one-dimensional harmonic oscillator states needed here are
$$
\psi_0(y_{{\rm pn}})=\left(\frac{\nu}{\pi}\right)^{\frac{1}{4}}~e^{-\frac{1}{2}\nu y_{{\rm pn}}^2}
$$
$$   
\psi_1(y_{{\rm pn}})=\left(\frac{\nu}{\pi}\right)^{\frac{1}{4}}~\sqrt{2 \nu}~y_{{\rm pn}}~e^{-\frac{1}{2}\nu y_{{\rm pn}}^2}~=~-\sqrt{\frac{2}{\nu}}~\frac{\partial}{\partial y_{{\rm pn}}}~\psi_0(y_{{\rm pn}})
$$
so that
$$
\left[\psi_1(y_{{\rm pn}})\psi_0(y_{{\rm pn}})\right]~=~-\sqrt{\frac{2}{\nu}}~\psi_0(y_{{\rm pn}})~\frac{\partial}{\partial y_{{\rm pn}}}~\psi_0(y_{{\rm pn}})~=~-\frac{1}{\sqrt{2 \nu}}~\frac{\partial}{\partial y_{{\rm pn}}}~\left[\psi_0(y_{{\rm pn}})\right]^2
$$
If we use this relation in Equation (\ref{Eqn3.6b})
\begin{eqnarray*}
\rho_{{{\rm g.s.}}\rightarrow{{\rm y}}}({\bf r'})~&=&~-\frac{e}{\sqrt{2 \nu}}~\int dx_{{\rm pn}}dy_{{\rm pn}}dz_{{\rm pn}}\left[\psi_0(x_{{\rm pn}})\right]^2~\frac{\partial}{\partial y_{{\rm pn}}}~\left[\psi_0(y_{{\rm pn}})\right]^2~\left[\psi_0(z_{{\rm pn}})\right]^2\\
&\times&G\left(\sqrt{\left(x'-f~x_{{\rm pn}}\right)^2+\left(y'-f~y_{{\rm pn}}\right)^2+\left(z'-f~z_{{\rm pn}}\right)^2}\right), 
\end{eqnarray*}
and integrate by parts, we get
\begin{eqnarray}
\rho_{{{\rm g.s.}}\rightarrow{{\rm y}}}({\bf r'})~&=&~+\frac{e}{\sqrt{2 \nu}}~\int dx_{{\rm pn}}dy_{{\rm pn}}dz_{{\rm pn}}\left[\psi_0(x_{{\rm pn}})\right]^2~\left[\psi_0(y_{{\rm pn}})\right]^2~\left[\psi_0(z_{{\rm pn}})\right]^2 \nonumber\\
&\times&\frac{\partial}{\partial y_{{\rm pn}}}~G\left(\sqrt{\left(x'-f~x_{{\rm pn}}\right)^2+\left(y'-f~y_{{\rm pn}}\right)^2+\left(z-f~z_{{\rm pn}}\right)^2}\right).\label{Eqn3.7}
\end{eqnarray} 
But
\begin{eqnarray*}
&&\frac{\partial}{\partial y_{{\rm pn}}}~G\left(\sqrt{\left(x'-f~x_{{\rm pn}}\right)^2+\left(y'-f~y_{{\rm pn}}\right)^2+\left(z'-f~z_{{\rm pn}}\right)^2}\right)\\
&~=~&-f ~\frac{\partial}{\partial y'}~G\left(\sqrt{\left(x'-f~x_{{\rm pn}}\right)^2+\left(y'-f~y_{{\rm pn}}\right)^2+\left(z'-f~z_{{\rm pn}}\right)^2}\right),
\end{eqnarray*}
so that Equation (\ref{Eqn3.7}) becomes 
\begin{eqnarray}  
\rho_{ {{\rm g.s.}}\rightarrow{{\rm y}} }({\bf r'})~&=&~-\frac{ef}{\sqrt{2 \nu}}~\frac{\partial}{\partial y'}~\int dx_{{\rm pn}}dy_{{\rm pn}}dz_{{\rm pn}}\left[\psi_0(x_{{\rm pn}})\right]^2~\left[\psi_0(y_{{\rm pn}})\right]^2~\left[\psi_0(z_{{\rm pn}})\right]^2 \nonumber\\
&\times&~G\left(\sqrt{\left(x'-f~x_{{\rm pn}}\right)^2+\left(y'-f~y_{{\rm pn}}\right)^2+\left(z'-f~z_{{\rm pn}}\right)^2}\right)\nonumber\\
&=&-\frac{f}{\sqrt{2 \nu}}~\frac{\partial}{\partial y'}\rho_{{\rm g.s.}}({\bf r'})~=~-\frac{f}{\sqrt{2 \nu}}~\frac{y'}{r'}\frac{\partial}{\partial r'}\rho_{{\rm g.s.}}(r')\nonumber\\
&~=~&-\frac{f}{\sqrt{2 \nu}}~\sin(\theta')\sin(\phi')\frac{\partial}{\partial r'}\rho_{{\rm g.s.}}(r')\label{Eqn3.8}.
\end{eqnarray}
Thus the radial shape of the transition charge density is obtained from the {\it radial derivative of the target ground state charge density}. This result depends upon the interpretation of the GDR as a harmonic oscillation of proton and neutron spheres relative to each other, but it makes no assumption about the radial charge dependence of the proton sphere. 

The occurrence of the derivative of the ground state radial density in expressions for transition densities is familiar from discussions of direct inelastic scattering to modes which are interpreted as small shape oscillations. An 
exhaustive discussion of this topic can be found, e.g., in Ch.14 of ref. \cite{Satchler}. These theories are based on expansions of the nuclear shape in powers of the small parameter describing the oscillation. We note that our derivation of Equation (\ref{Eqn3.8}) is not based on an expansion in powers of a small parameter. However, our use of harmonic oscillator wave functions in the derivation of Equation (\ref{Eqn3.8}) depended upon our assumption that the GDR potential is harmonic, which will generally be true only if the deviation from equilibrium is small.

The size parameter $\nu$ used in Equations (\ref{Eqn3.6a}) through (\ref{Eqn3.8}) is given by
$\nu=\mu \omega/\hbar$, where $\mu$ is the reduced mass for the neutron and proton spheres, and $\omega$ is the GDR oscillation frequency. We will use the generic formula \cite{BM}
\beq
\hbar \omega=79~A_{{\rm T}}^{-\frac{1}{3}},
\label{Eqn3.9}
\eeq
where $A_{{\rm T}}$ is the target mass number.

\subsection{The transition matrix element}
Some of the basic formulae of the semi-classical approach to RCE are presented in the Appendix.

The time-dependent target transition matrix element is
\beq
V_{{\rm g.s.}\rightarrow{\rm y}}(t',b)=\int d^3r'\left[~\rho_{{\rm g.s.}\rightarrow {\rm y}} ({\bf r'}) \phi_{_{{\rm P}}}'({\bf r'},t')-\frac{1}{c}{\bf j}_{{{\rm g.s.}}\rightarrow{{\rm y}} }({\bf r'})\cdot{\bf A}_{_{{\rm P}}}'(\bf{ r'},t')~\right].
\label{Eqn3.10}
\eeq
According to Equations (\ref{Eqn3.4}) and (\ref{Eqn3.10}), only the $z$ component of the transition current density enters into the matrix element. This is zero for the matrix element connecting the zero-phonon state to the state with one phonon of oscillation in the ${\bf {\hat y}}$ direction. Thus, for this particular matrix element, we need only be concerned with the $\rho-\phi$ term in (\ref{Eqn3.10}). 

The partition of the scalar potential into LW and overlap terms carries over to the transition matrix element:
\bes
\bea
V_{{\rm g.s.}\rightarrow{\rm y}}(t',b)~&=&~V_{{\rm LW}}(t')~+~V_{{\rm overlap}}(t')\\
V_{{\rm LW}}(t',b)~&=&~\int d^3r'\frac{\gamma Z_{_{\rm P}}e}{\sqrt{x'^2+(y'-b)^2+\gamma^2(z'-vt')^2}}\times\rho_{{{\rm g.s.}}\rightarrow{{\rm y}} }({\bf r'})\label{Eqn3.11a}\\
V_{{\rm overlap}}(t',b)~&=&~4\pi\gamma\int d^3r'\int_{s=r}^\infty (s-\frac{s^2}{r})\rho_{_{\rm P}} (s)ds\times\rho_{{{\rm g.s.}}\rightarrow{{\rm y}} }({\bf r'})\label{Eqn3.11b}\\
&~&{\rm~~~~~with~~}r\equiv\sqrt{x'^2+(y'-b)^2+\gamma^2(z'-vt')^2}\nonumber 
\eea
\ees
For the purposes of calculation, it is convenient to define a function $F(r)$ by
\beq
F(r)~\equiv~4 \pi \gamma \int_{s=r}^\infty s(s-r)\rho_{_{\rm P}} (s)ds.
\label{Eqn3.12}
\eeq
Then $V_{{\rm LW}}(t')$ and $V_{{\rm overlap}}(t')$ can be written more compactly as
\bes
\bea
V_{{\rm LW}}(t')&=&\int d^3 r'\frac{\rho_{{{\rm g.s.}}\rightarrow{{\rm y}} }({\bf r'})}{r}F(0)
\label{Eqn3.13a}\\
V_{{\rm overlap}}(t')&=&-\int d^3 r'\frac{\rho_{{{\rm g.s.}}\rightarrow{{\rm y}} }({\bf r'})}{r}F(r)
\label{Eqn3.13b}\\
&~&{\rm~~~~~with~~}r\equiv\sqrt{x'^2+(y'-b)^2+\gamma^2(z'-vt')^2}\nonumber 
\eea
\ees
Once $\rho_{_{\rm P}} (s)$, the radial form of the projectile proton density distribution, has been chosen, $F(r)$ is calculated using Equation(\ref{Eqn3.12}), and it is then used in the numerical evaluation of the integrals (\ref{Eqn3.13a},\ref{Eqn3.13b}).

\subsection{Cross-sections}
Up to bombarding energies of about 5 GeV per nucleon, first-order time-dependent perturbation theory accounts for almost all of the population of the state with one-phonon of oscillation in the ${\bf {\hat y}}$ direction (see, e.g., 
\cite{BZ68}). This leads to
$$
P_{{\rm g.s.} \rightarrow {\rm y}}(b)=\left|~\int_{-\infty}^\infty \frac{dt'}{\hbar}e^{i \omega t'}V_{{\rm g.s.}\rightarrow{\rm y}}(t',b)~\right|^2\equiv \left|{\tilde V}_{{\rm g.s.}\rightarrow {\rm y}}(\omega, b)\right|^2,
$$
where ${\tilde V}_{{\rm g.s.} \rightarrow {\rm y}}(\omega, b)$ is the ``on-shell" Fourier component of $V_{{\rm g.s} \rightarrow {\rm y}}(t',b)$ (see Appendix). Using Equation (\ref{Eqn1.2}), we can write\footnote{These matrix elements are real.}
$$
\sigma=\int b db \left| {\tilde V}_{{\rm g.s.} \rightarrow {\rm y}}(\omega,b) \right|^2X(b)=\int b db \left| 
{\tilde V}_{{\rm LW}}(\omega,b)+ {\tilde V}_{{\rm overlap}}(\omega,b)\right|^2X(b)
$$
\bea
~&=&\int b db \left({\tilde V}_{{\rm LW}} (\omega,b)\right)^2X(b)~+~\int b db ~{\tilde V}_{{\rm overlap}}(\omega,b)\times\left(2 {\tilde V}_{{\rm LW}} (\omega,b)+{\tilde V}_{{\rm overlap}} (\omega,b)\right)X(b)\nonumber \\
~&=&~~~~~~~~~~~~\sigma_{{\rm LW}}~~~~~~~~~~~~~~+~~~~~~~~~~~~~~~\sigma_{{\rm overlap}}.
\label{Eqn3.14}
\eea 
Here $\sigma_{{\rm LW}}$ is the cross-section that would have been calculated had the Lienard-Wiechert potential been used {\it everywhere}, even in the overlap region, and $\sigma_{{\rm overlap}}$ is the correction that must be applied due to the inadequacy of Lienard-Wiechert potential.

Application of the 
Winther-Alder \cite{WA} general formula, Equation (\ref{EqnA.1}), to ${\tilde V}_{{\rm LW}}(\omega,b)$ gives 
\beq
{\tilde V}_{{\rm LW}}(\omega,b)=- \frac{\pi c Z_{{\rm P}}ef}{\hbar \gamma v^2}\sqrt{\frac{32}{\nu}} K_1\left(\frac{\omega b}{\gamma v}\right) \int_0^\infty r'^2 dr' j_1\left(\frac{\omega}{c}r'\right)\frac{\partial \rho_{{\rm g.s.}}(r')}{\partial r'}.
\label{Eqn3.15}
\eeq 
We can use this in Equation (\ref{Eqn3.14}), but ${\tilde V}_{{\rm overlap}}(\omega,b)$ must be evaluated numerically, as must the $b$ integral in $\sigma_{{\rm overlap}}$. The latter is simplified by the fact that the effective $b$ range is finite. It is limited from below by the vanishing of $X(b)$, and is limited from above by the vanishing of ${\tilde V}_{{\rm overlap}}(\omega,b)$, since large $b$ implies small overlap.

Table 2 shows a comparison of $\sigma_{{\rm LW}}$ and $\sigma_{{\rm overlap}}$ for $^{16}$O and $^{208}$Pb projectiles bombarding $^{74}$Ge and $^{202}$Hg targets at 2 GeV per nucleon. It is seen that $\sigma_{{\rm LW}}$ is very much larger when $^{208}$Pb is the projectile, and when $^{202}$Hg is the target. This is because {\it all} the projectile and target charge contribute to the LW cross-section. However the amount of overlap charge is approximately the same in all four cases, so that the four overlap corrections are of the same order of magnitude. Another feature that surpresses the overlap correction in the heavier systems is their greater neutron/proton ratio. The extra neutrons contribute to the absorption, but not to the electromagnetic interaction.

Although $\sigma_{{\rm overlap}}$ is a greater fraction of $\sigma_{{\rm LW}}$ in lighter systems than in heavier sstems, we can see from Table 2 that $\sigma_{{\rm overlap}}$ is {\it always} very small compared to $\sigma_{{\rm LW}}$. Thus we can safely ignore $\sigma_{{\rm overlap}}$, and use $\sigma_{{\rm LW}}$ alone to account for experimental data.

\section{Projectile and target dependence of $b_{{\rm min}}$}
Suppose that $X(b)$ makes a sharp transition from 0 to 1 as $b$ crosses a particular value $b_{{\rm min}}$. Then Equation (\ref{Eqn3.14}) yields
\beq
\sigma_{{\rm LW}}=\int_{b=b_{{\rm min}}}^\infty b db \left({\tilde V}_{{\rm LW}} (\omega,b)\right)^2
\label{Eqn4.1}
\eeq
Application of Equations (\ref{EqnA.1}) through (\ref{EqnA.4})) gives
\bea
\sigma_{{\rm LW}}&=&\pi b_{{\rm min}}^2\left[~\left( K_2\left(\frac{\omega b_{{\rm min}} }{\gamma v}\right) \right)^2-\left(K_1\left(\frac{\omega b_{{\rm min}} }{\gamma v}\right) \right)^2-2\left(\frac{\gamma v}{\omega b_{{\rm min}} }\right)^2 K_2\left(\frac{\omega b_{{\rm min}} }{\gamma v}\right) K_1\left(\frac{\omega b_{{\rm min}} }{\gamma v}\right)~\right]\nonumber \\
~&\times&\left(\frac{\pi c Z_{{\rm P}}ef}{\hbar \gamma v^2}\sqrt{\frac{32}{\nu}} \int_0^\infty r'^2 dr' j_1\left(\frac{\omega}{c}r'\right)\frac{\partial \rho_{{\rm g.s.}}(r')}{\partial r'}\right)^2,  
\label{Eqn4.2}
\eea
It would be advantageous to have a prescription for $b_{{\rm min}}$ such that the entire expression (\ref{Eqn3.14}) for the cross-section would be given by the explicit formula Equation(\ref{Eqn4.2}). Equivalently, we seek a formula for $b_{{\rm min}}$ that satisfies\footnote{Bertulani {\it et al} \cite{BCH} use the symbol $b_{{\rm sharp}}$ to represent this quantity.} 
$$
\int_{b=0}^\infty b db~\left[~{\tilde V}_{{\rm LW}}(\omega,b)^2~\right]~X(b)=\int_{b=b_{{\rm min}}}^\infty b db~\left[~{\tilde V}_{{\rm LW}}(\omega,b)^2~\right].
$$
Since all the $b$ dependence of ${\tilde V}_{{\rm LW}}(\omega,b)$ is contained in $K_1\left(\frac{\omega b_{{\rm min}} }{\gamma v}\right)$, we can write this equation more explicitly as
\beq
\int_{b=0}^\infty b db~\left(K_1\left(\frac{\omega b }{\gamma v}\right) \right)^2~X(b)=\int_{b=b_{{\rm min}}}^\infty b db~\left(K_1\left(\frac{\omega b }{\gamma v}\right) \right)^2
\label{Eqn4.3}
\eeq
We could use $b_{{\rm min}}$ defined in this way in Equation (\ref {Eqn4.2}), thus incorporating the effect of nuclear interactions during grazing collisions of the projectile and target.

Since the left-hand side of Equation (\ref{Eqn4.3}) must be evaluated numerically, we cannot produce a closed formula for $b_{{\rm min}}$. However we can numerically evaluate Equation (\ref{Eqn4.3}) for a variety of projectiles, targets, and bombarding energies, and then look for regularities that could guide our choice of $b_{{\rm min}}$ in any particular situation. To carry out this program, we have selected as projectiles the nuclei $^{16}$O, $^{40}$Ca, $^{120}$Sn, and $^{208}$Pb, and as targets the nuclei $^{16}$O, $^{32}$S, $^{52}$Cr, $^{74}$Ge, $^{90}$Zr, $^{114}$Cd, $^{138}$Ba, $^{158}$Gd, $^{180}$Hf, and $^{202}$Hg. The GDR oscillation frequencies were given by Equation (\ref{Eqn3.9}). For each projectile-target combination, we have varied the bombarding energy from 1 to 5 GeV per nucleon, in steps 1 GeV per nucleon, and used Equation (\ref{Eqn4.3}) to calculate the appropriate value of $b_{{\rm min}}$.

In the limit of very large nucleon-nucleon interaction cross-sections, we would expect $b_{{\rm min}}$ to be of the order of $R_{{\rm P}}+R_{{\rm T}}$, or perhaps somewhat larger if the finite range of the nucleon-nucleon interaction is included. This would lead to an expression for $b_{{\rm min}}$ proportional to $A_{{\rm P}}^{1/3}+A_{{\rm T}}^{1/3}$. However, a more realistic nucleon-nucleon interaction would allow the projectile and target densities to overlap slightly without nuclear interaction, which would lead to a smaller value of $b_{{\rm min}}$. If the projectile and/or the target have large radii, even a small amount of penetration implies a large overlap volume, and thus a large nuclear interaction probability. Thus the downward correction to an $A_{{\rm P}}^{1/3}+A_{{\rm T}}^{1/3}$ term would be expected to decrease as the size of the colliding nuclei increases. The simplest way to incorporate these trends into a formula is to seek parameters $\lambda, \mu$ such that
\beq
b_{{\rm min}}\sim \lambda \left(A_{{\rm P}}^{1/3}+A_{{\rm T}}^{1/3}\right)~-~\mu \left(A_{{\rm P}}^{-1/3}+A_{{\rm T}}^{-1/3}\right).
\label{Eqn4.4}
\eeq
This is the form used by Benesh {\it et al} \cite{BCV} in their analysis of the total reaction cross-section for colliding nuclei. We have chosen $\lambda$ and $\mu$ to produce the best fit, in a least-squares sense to the 200 (4 $\times$ 10 $\times$ 5) projectile-target-energy combinations for which we have numerically calculated $b_{{\rm min}}$ using Equation (\ref{Eqn4.3}). The result is
$$
\lambda=1.3115 {\rm fm},~~~~~\mu=1.0509 {\rm fm}.
$$ 
If these paramaters are used in Equation (\ref{Eqn4.4}), the calculated values of $b_{{\rm min}}$ are reproduced with an r.m.s. error of 0.0248 fm per point. A graphical comparison is shown in Figure 3. The continuous lines are plots of Equation (\ref{Eqn4.4}) for our four projectiles, and the plotted points refer to our numerical calculations of $b_{{\rm min}}$ for our ten targets at bombarding energies of 1 and 5 GeV per nucleon. It is seen that Equation (\ref{Eqn4.4}), with the parameters given above, provides an excellent representation of the $b_{{\rm min}}$ values numerically calculated from Equation (\ref{Eqn4.3}).

Another study \cite{KGP} of nucleus-nucleus reaction cross-sections adopted a form equivalent to
\beq
b_{{\rm min}}\sim \lambda \left(A_{{\rm P}}^{1/3}+A_{{\rm T}}^{1/3}\right)~+~\mu \frac{A_{{\rm P}}^{1/3}\times A_{{\rm T}}^{1/3}}{A_{{\rm P}}^{1/3}+A_{{\rm T}}^{1/3}}-c
\label{Eqn4.5}
\eeq
This equation contains three adjustable parameters, $\lambda, \mu$ and $c$. A least-square-deviation fit to our 200 values of $b_{{\rm min}}$ yields
$$
\lambda=1.3338 {\rm fm},~~~~~\mu=0.1652 {\rm fm},~~~~~c=1.0667 {\rm fm},
$$
with an r.m.s error of 0.0382 fm per point. We see that the form (\ref{Eqn4.5}) with three free parameters does not produce as good an overall fit as the form (\ref{Eqn4.4}), which has only two free parameters. We conclude that Equation (\ref{Eqn4.4}) gives the most effective and economical representation of the dependence of $b_{{\rm min}}$ on projectile and target mass numbers. 

The $b_{{\rm min}}$ values given by Equations (\ref{Eqn4.4}) or (\ref{Eqn4.5}) yield very nearly the same CEX cross-sections when substituted into Equation (\ref{Eqn4.2}). The differences are less than 1\% for all cases except for 1 GeV per nucleon $^{16}$O projectiles on a $^{16}$O target, where the difference is 2\%. These differences are probably small compared to the errors associated with the use of first-order time-dependent perturbation theory. 

\section{Discussion}

All the calculations presented so far have referred to $\mu=1$ transitions, {\it i.e.} transitions in which the transfer of the ${\bf {\hat z}}$ component of angular momentum is $\pm \hbar$. These are the dominant transitions in RCE. Indeed, in the Weizs\"{a}cker-Williams \cite{WeiWil} approach to RCE, these are the {\it only} transitions considered. Nevertheless, transtions with $\mu \neq 1$ are possible, and it is of some interest to know how a change in $\mu$ would affect $b_{{\rm min}}$. This can be answered simply by replacing $K_1$ in Equation (\ref{Eqn4.3}) by $K_\mu$. In the vicinity of $b\sim b_{{\rm min}}$, the argument of $K_\mu\left(\frac{\omega b_{{\rm min}} }{\gamma v}\right)$ is small, which implies that $K_\mu$ is proportional to $b^{-\mu}$. Thus $K_\mu$ falls more sharply with increasing $b$ as $\mu$ increases, and this has the consequence that $b_{{\rm min}}$ calculated from Equation (\ref{Eqn4.3}) will decrease as $\mu$ increases. Another way to reach this conclusion is to think about the $\mu$ component of the transition charge density. As $\mu$ increases, the centrifugal potential will keep the transition charge density farther away from the ${\bf {\hat z}}$ axis. Thus absorption, which occurs close to the ${\bf {\hat z}}$ axis, will have less effect on high $\mu$ transitions. Since it is absorption that gives rise to a minimum effective value of $b$, less absorption will mean a smaller value of $b_{{\rm min}}$.

We have done calculations for $\mu=2$, and found a decrease in $b_{{\rm min}}$, compared to the $\mu=1$ values presented in the last section, by about 0.05 fm when the projectile is $^{208}$Pb. For the range of targets and bombarding energies we have studied, this decrease in $b_{{\rm min}}$ is equivalent to an increase in CEX cross-section of about 0.7\% to 1.5\%. This difference is small enough to be ignored in practical calculations. If the projectile is $^{16}$O, the corresponding error in $b_{{\rm min}}$ is about 0.1 fm, which is equivalent to errors of about 1.5\% to 7\% in CEX cross-sections. Thus, for lighter projectiles, and in situations in which $\mu>1$ transitions are expected to be important, the values of $b_{{\rm min}}$ calculated with Equations (\ref{Eqn4.4}), (\ref{Eqn4.5}) should be interpreted only as upper limits.

To do a calculation that includes the effect of absorption accurately when several values of $\mu$ are important, it is necessary to give up the picture implied by Equation (\ref{Eqn1.2}), in which absorptive processes occur independently of the Coulomb excitation. Rather, the coupled equations that are used to calculate the transition amplitude would have to include absorptive processes along with electromagnetic processes. This would be analagous to the way that optical model analyses of inelastic scattering employ an imaginary potential to simulate absorption into other channels in the calculation of the inelastic scattering amplitude.

\appendix
\section{Some basic formulae of the semi-classical approach to relativistic Coulomb excitation.}
The Fourier transform of the matrix element for the transfer of angular momenta $(\lambda,\mu)$ in the target transition $\phi_\alpha\rightarrow\phi_\beta$, due to the time-dependent electromagnetic field of a spherically symmetric projectile moving with speed $v$ along a trajectory with impact parameter $b$ is
\bea
V_{\beta \alpha}(\omega,b)&\equiv&\int_{-\infty}^\infty \frac{dt'}{\hbar}V_{\beta \alpha}(t',b)
\nonumber \\
~&=&\frac{2 Z_{{\rm P}}e}{\hbar v}e^{-i\phi_b}\left[{\cal G}_{\lambda, \mu}\int d^3r'\left( \rho_{\beta \alpha}({\bf r'})-\frac{v}{c^2}{\bf {\hat z}\cdot j}_{\beta \alpha}({\bf r'})\right)j_\lambda\left(\frac{|\omega|}{c}r'\right)Y^\lambda_\mu({\hat r'})\right]\nonumber \\
~&\times&~K_\mu\left(\frac{|\omega| b}{\gamma v}\right).
\label{EqnA.1}
\eea
Here $\rho_{\beta \alpha}({\bf r'})$ and ${\bf j}_{\beta \alpha}({\bf r'})$ are the target charge and current transition charge densities for the states $\phi_\alpha\rightarrow\phi_\beta$, and $K_\mu$ is a modified Bessel function. The coefficients ${\cal G}_{\lambda, \mu}$ are defined by
\bea
{\cal G}_{\lambda, \mu}&\equiv&\frac{i^{\lambda+\mu}}{(2 \gamma)^\mu}\left(\frac{|\omega|}{\omega}\right)^{\lambda - \mu}\sqrt{4 \pi (2 \lambda+1)(\lambda-\mu)!(\lambda+\mu)!}\nonumber \\
&\times&\sum_n\frac {1}{(2 \gamma)^{2n}(n+\mu)!n!(\lambda-\mu-2n)!}
\label{EqnA.2}
\eea
This expression assumes there is no overlap between projectile and target charge. The {\it time reversal} phase convention is used, so that the spherical harmonics $Y^\ell_m$ has an extra factor of $i^{\ell}$ compared to a {\it Condon-Shortley} spherical harmonic.

If it is assumed that no contribution to RCE occurs from $b<b_{{\rm min}}$ and no nuclear interactions occur for $b>b_{{\rm min}}$, then the first-order time-dependent perturbation theory approximation for the cross-section for the RCE population of a state $\phi^\lambda_\mu$ in an even-even nucleus is
\beq
\sigma = 2\pi \int_{b=b_{{\rm min}}}^\infty b db\left|V_{\beta \alpha}(\omega,b)\right|^2
\label{EqnA.3}
\eeq
The value of $\omega$ to be used here is $\frac{1}{\hbar}$ times the excitation energy of $\phi^\lambda_\mu$. This is sometimes referred to as the ``on-shell" $\omega$ value. For the calculations in this paper, the excited state corresponds to a one-phonon excitation of the GDR, and thus the on-shell value of $\omega$ is given by Equation (\ref{Eqn3.9}). Because all the $b$ dependence of $V_{\beta \alpha}(\omega,b)$ is contained in the factor $K_\mu\left(\frac{|\omega| b}{\gamma v}\right)$, the $b$ integral in Equation ({EqnA.3}) can be performed exactly, with the help of 
\beq
\int_\xi^\infty \left( K_\mu(x)\right)^2 x dx=\frac{\xi^2}{2}\left[\left( K_{\mu+1}(\xi)\right)^2-\left( K_\mu(\xi)\right)^2-\frac{2 \mu}{\xi}\left(K_{\mu+1}(\xi)K_\mu(\xi)\right)\right]
\label{EqnA.4}
\eeq

\newpage

\centerline{Figure Captions}
\medskip
Figure 1. The reactivity of a projectile nucleon is represented by a grey disk (seen edgewise) located at ${\bf r_{{\rm P}}}$ relative to the projectile center. Points on the disk are labelled by ${\bf s}$ relative to the center of the disk, and by ${\bf r_{{\rm P}}}$ relative to the target center.

Figure 2. Goldhaber-Teller picture of the GDR, with proton and neutron spheres oscillating relative to each other. The oscillation variable is ${\bf r_{{\rm PN}}}$, the vector connecting the centers of the two spheres.

Figure 3. The circles correspond to $b_{{\rm min}}$ values calculated using Equation (\ref{Eqn4.3}), when the bombarding energy is 1 GeV per nucleon. The crosses are the same, but for a bombarding energy of 5 GeV per nucleon. The continuous lines were calculated using Equation (\ref{Eqn4.4}), with $\lambda=1.3115$ fm and $\mu=1.0509$ fm.
\newpage
\begin{table}
\begin{center}
\begin{tabular}{|c|c|c|c|c|}
\hline
Bombarding energy & $\sigma_{{\rm ZZ}}$ & $B_{{\rm ZZ}}$ & $\sigma_{{\rm ZN}}$ & $B_{{\rm ZN}}$ \\
(GeV per nucleon) & (mb) & (GeV/c)$^{-2}$ & (mb) & (GeV/c)$^{-2}$ \\ \hline
1 & 47.849 & 5.814 & 40.221 & 4.1 \\ \hline
2 & 45.024 & 6.349 & 42.973 & 5.904 \\ \hline
3 & 42.493 & 6.847 & 42.520 & 6.645 \\ \hline
4 & 41.307 & 7.280 & 42.124 & 7.384 \\ \hline
5 & 40.809 & 7.737 & 42.567 & 8.069 \\ \hline
\end{tabular}
\caption[smallcaption]{Parameters to determine the spin-averaged nucleon-nucleon scattering amplitudes, interpolated and extrapolated from figures given by Igo \cite{Igo}.}
\end{center}
\end{table}
\renewcommand{\arraystretch}{1.5}
\begin{table}
\begin{center}
\begin{tabular}{|c|c|c|c|c|}
\hline
~& \multicolumn{2}{c}{$^{74}$Ge target} \vline & \multicolumn{2}{c}{$^{202}$Hg target} \vline\\ \hline 
Projectile & $\sigma_{{\rm LW}}$ (barns) & $\sigma_{{\rm overlap}}$ (barns) & $\sigma_{{\rm LW}}$ (barns) & $\sigma_{{\rm overlap}}$ (barns) \\ \hline
$^{16}$O & $1.9\times 10^{-2}$ & $-9.3\times 10^{-5}$ & $7.3\times 10^{-2}$ & $-9.9\times 10^{-5}$ \\ \hline 
$^{208}$Pb & 1.3 & $-3.5\times 10^{-4}$ & 5.8 & $-3.5\times 10^{-4}$ \\ \hline
\end{tabular}
\caption[smallcaption]{$\sigma_{{\rm LW}}$ and $\sigma_{{\rm overlap}}$, defined in Equation (\ref{Eqn3.14}), for $^{16}$O and $^{208}$Pb projectiles bombarding $^{74}$Ge and $^{202}$Hg targets, at 2 GeV per nucleon.}
\end{center}
\end{table}

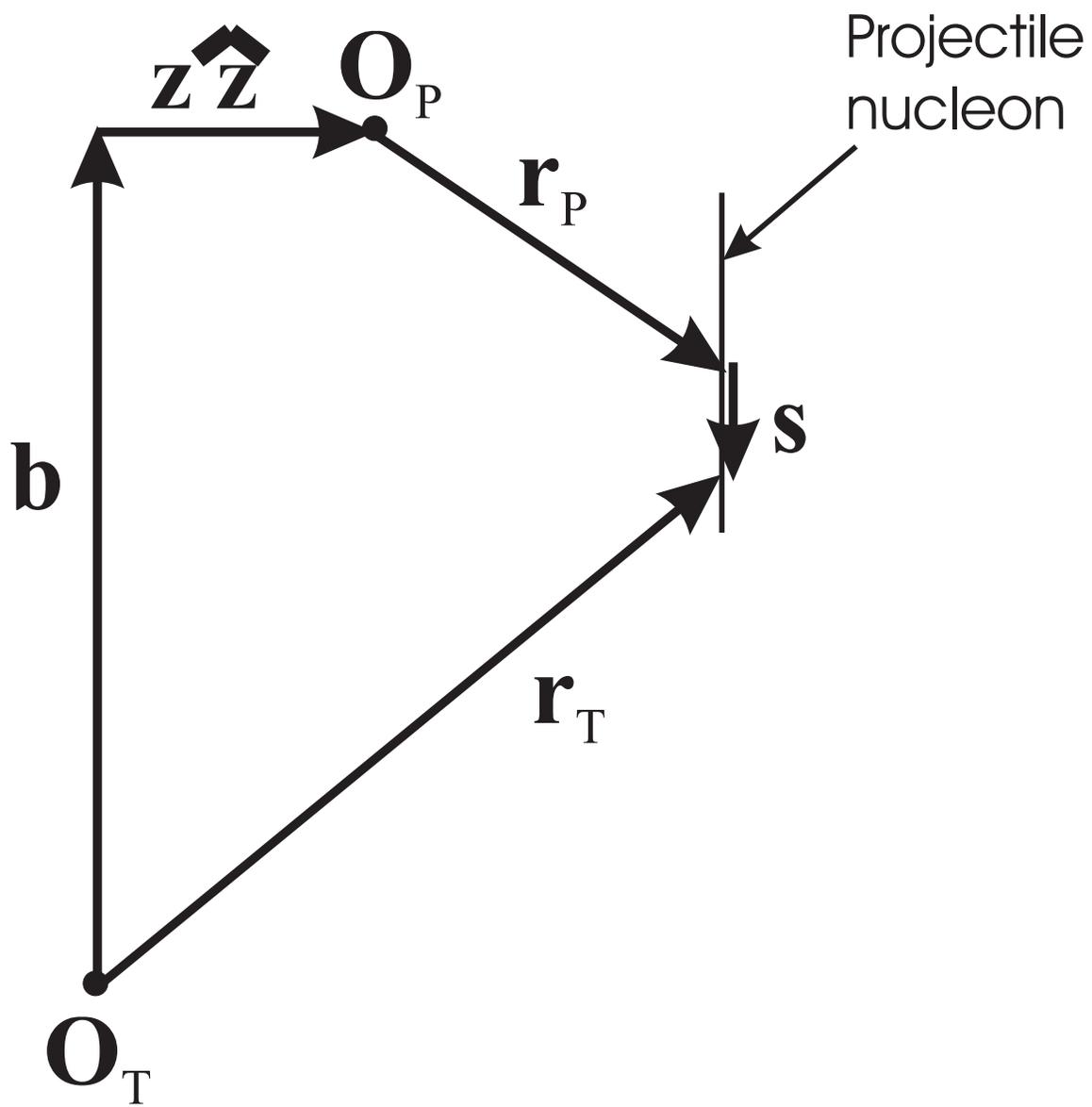
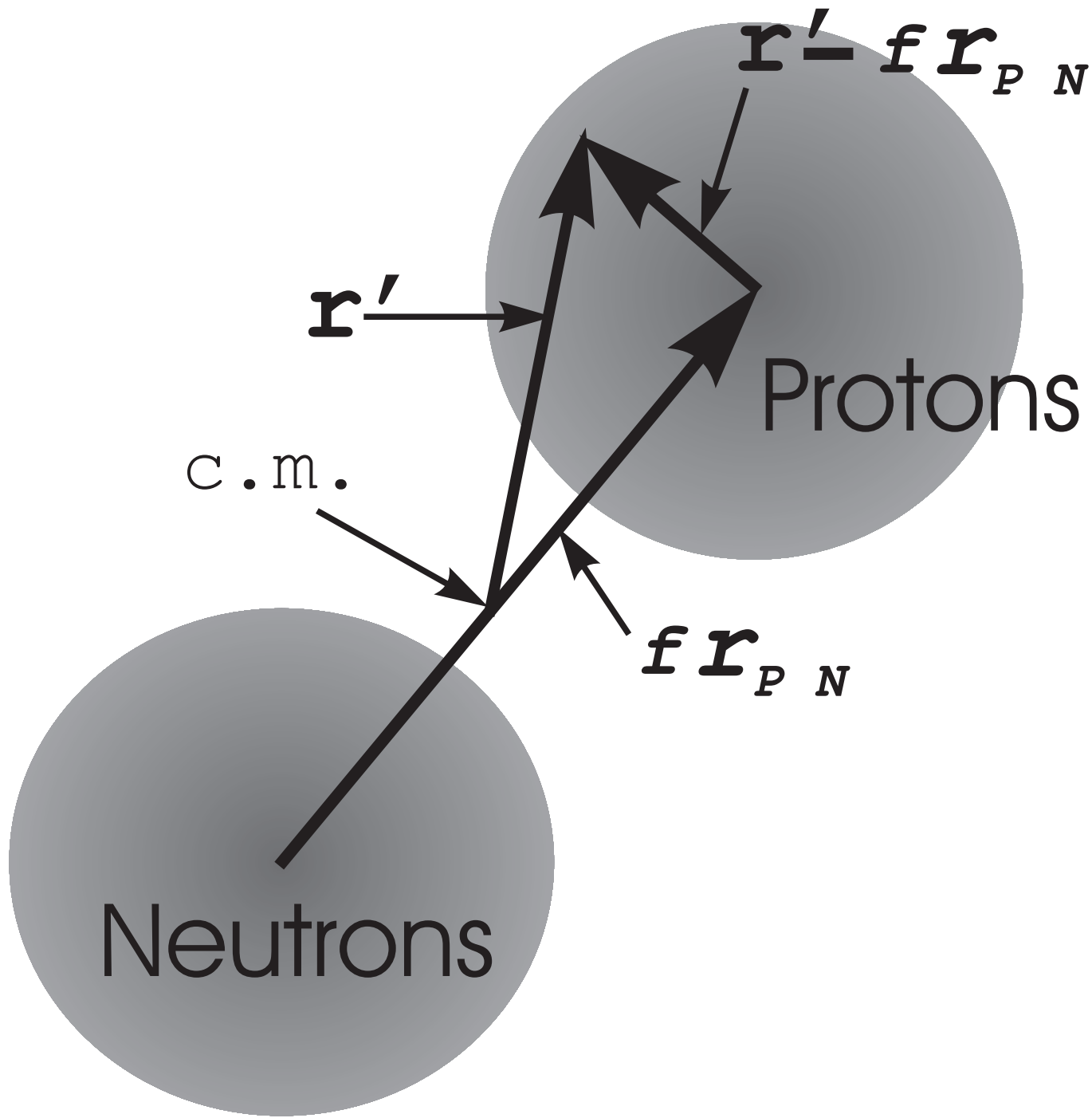
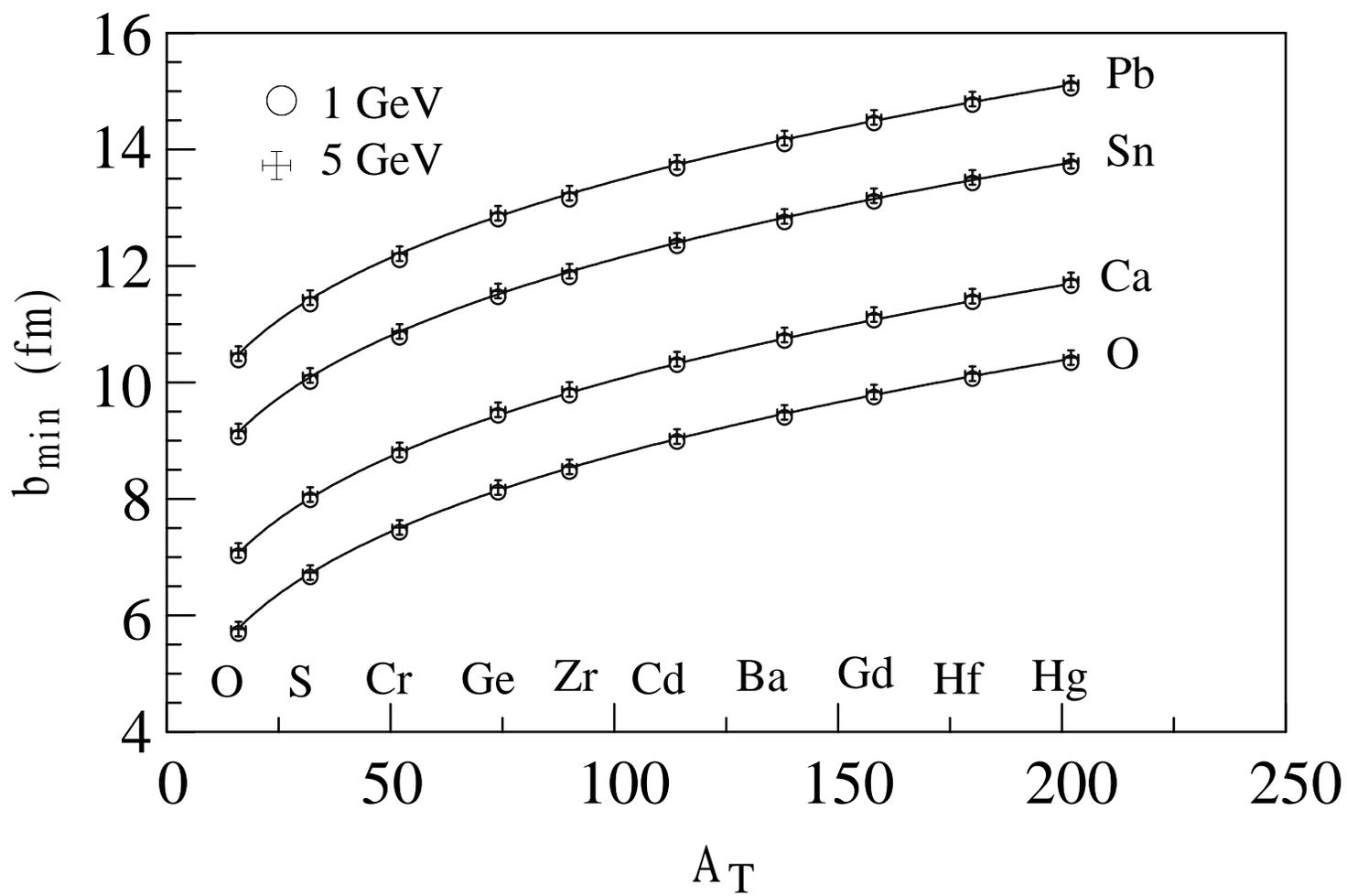
\end{document}